\documentstyle[amsmath,amssymb,graphicx]{article}

\def\be{\begin{eqnarray}}
\def\ee{\end{eqnarray}}
\def\nn{\nonumber}

\def\p{\partial}

\hoffset= - 1.0in         

\title{{\bf Matrix Model Conjecture for Exact BS Periods
and Nekrasov Functions} \vspace{.2cm}}
\author{{\bf A.Mironov}\footnote{ {\small {\it
Lebedev Physics Institute} and {\it ITEP, Moscow, Russia}};
mironov@itep.ru; mironov@lpi.ru}, {\bf A.Morozov}\thanks{{\small
{\it ITEP, Moscow, Russia}}; morozov@itep.ru} \ and {\bf
Sh.Shakirov}\thanks{{\small {\it ITEP, Moscow, Russia} and
{\it MIPT, Dolgoprudny, Russia}};
shakirov@itep.ru}\date{ }}

\begin{document}

\maketitle

\vspace{-6.0cm}

\begin{center}
\hfill FIAN/TD-28/09\\
\hfill ITEP/TH-69/09\\
\end{center}

\vspace{4cm}

\begin{abstract}
We give a concise summary of the impressive recent development
unifying a number of different fundamental subjects.
The quiver Nekrasov functions (generalized hypergeometric series)
form a full basis for all conformal blocks of the Virasoro
algebra and are sufficient to provide the same for some (special)
conformal blocks of W-algebras.
They can be described in terms of Seiberg-Witten theory,
with the SW differential given by the 1-point resolvent
in the DV phase of the quiver (discrete or conformal) matrix model
($\beta$-ensemble),
$dS = ydz + O(\varepsilon^2) =
\sum_p \varepsilon^{2p}\rho_\beta^{(\!p\,|1)}\!(z)$,
where $\varepsilon$ and $\beta$ are related to the LNS
parameters $\epsilon_{1}$ and $\epsilon_2$.
This provides explicit formulas for conformal blocks in terms of analytically continued
contour integrals and resolves the old puzzle of the free-field description of
generic conformal blocks through the Dotsenko-Fateev integrals.
Most important, this completes the GKMMM description of SW
theory in terms of integrability theory with the help of
exact BS integrals,
and provides an extended manifestation of the basic principle which states
that the effective actions are the tau-functions of
integrable hierarchies.
\end{abstract}

\bigskip

\bigskip

{\bf 1. Introduction.}
A renewed interest to SW theory \cite{SW1}-\cite{SWl} and to Nekrasov functions
\cite{Nf1}-\cite{Nfl} is caused by the AGT conjecture \cite{AGT1}-\cite{AGTl},
which connects the subject to the classical field of conformal field theory
\cite{CFT1}-\cite{CFTl}.
Today it is clear that a new unification of the fundamental importance emerges,
bringing together at a principally new level the
CFT, the theory of loop algebras, SW theory, quantization theory, Baxter equations,
DV phase of matrix models, loop equations, the theory of hypergeometric functions,
symmetric groups, Hurwitz theory, Kontsevich models and modern combinatorics.
This unification is capable to resolve a number of long-standing problems
in each of the fields.
The goal of this paper is to briefly summarize our knowledge about this
emerging pattern, which is scattered and expressed in length in a number of fresh
\cite{AGT1}-\cite{AGTl} and older \cite{Wirc}-\cite{KS} papers.
The main emphasize will be put on the description of Nekrasov functions
in terms of SW theory, where the exact prepotential is expressed through
the exact Bohr-Sommerfeld integrals, and the integrand is provided by the
1-point function of conformal matrix model (or quiver $\beta$-ensemble)
in the Dijkgraaf-Vafa (DV) phase \cite{DV1}-\cite{DVl}.
This conjecture, explicitly formulated in \cite{DVagt} and further investigated
\cite{Ito,Egu}, makes the picture complete,
resolves the remaining uncertainties (about the shape of the second deformation)
in \cite{NS,MMBS1}
and finalizes the program \cite{GKMMM}
to reformulate SW theory of \cite{SW1,SW2} in terms of the BS integrals and
underlying integrable systems.
We do not discuss long formulas, checks and even evidence in favor of all these
conjectures: all calculations in these fields remain long and tedious,
and most statements still need to be checked and proved, however, the entire
picture is starting to get relatively clear.

\bigskip

{\bf 2. Nekrasov functions from BS/SW periods.}
The central object of emerging unification is the exact SW-Nekrasov prepotential
${\cal F}(\vec a|\epsilon_1,\epsilon_2) = \epsilon_1\epsilon_2\log Z_{Nek}$,
which now has a number of different interpretations:
\be
(A) & Z_{Nek} = {\rm sum\ over\ chains\ of\ Young\ diagrams} =
{\rm generalized\ hypergeometric\ series}
\nn\\
(B) & Z_{Nek} = {\rm sum\ over\ partitions\ with\ Plancherel\ like\ measure} =
{\rm discretized\ matrix\ model}\nn \\
(C) & Z_{Nek} = {\cal B} = {\rm conformal\ block,\ depending\ on\ a\ number\
of\ external\ and\ internal\ dimensions} \nn \\
(D) &
Z_{Nek}
= {\rm partition\ function\ of\ conformal}\ \beta-{\rm ensemble\ in\ DV\ phase}
= {\rm generalized\ Dotsenko-Fateev\ integrals}
\nn \\
(E) & Z_{Nek} = {\rm solution\ to\ consistent\ set\ of\ SW\ equations} \nn
\ee
\vspace{-0.8cm}
\be
a_I=\oint_{A_I} dS_{\epsilon_1,\epsilon_2}, \nn \\
\frac{\p{\cal F}}{\p a_I}=\oint_{B_I} dS_{\epsilon_1,\epsilon_2}, \nn
\ee
with deformed SW differential
\be
\boxed{
dS_{\epsilon_1,\epsilon_2}
= \Big(y + O(\varepsilon^2)\Big)dz = \sum_p \varepsilon^{2p}\rho_\beta^{(p|1)}(z)
}
\label{dS}
\ee

$\bullet$
$Z_{Nek}$ is the generalized hypergeometric series \cite{MMnf},
a sum over chains of Young diagrams (integer partitions),
introduced by N.Nekrasov \cite{Nf1} as expansion of the LNS multiple
contour integrals \cite{LNS} originally obtained by the localization
(Duistermaat-Heckman) technique \cite{DHth} from regularized integrals
over the ADHM moduli spaces \cite{ADHM} of instantons \cite{BPTS}
(for their original relation to SW theory and detailed references see
\cite{Khorev}).

$\bullet$
Discrete sums over integer "eigenvalues" are immediately associated
with the Nekrasov expansion of $Z_{Nek}$ into Young diagrams.
They are examples of the character expansions of $\tau$-functions \cite{unint}
and have deep and far-going relation to combinatorics of symmetric groups \cite{sygr},
to Hurwitz theory \cite{Hur} with the cut-and-join operators \cite{cajop},
and finally to the matrix models of Kontsevich type \cite{Kon}.

$\bullet$
The fact that $Z_{Nek}$ is the same as the conformal block is the
celebrated AGT conjecture \cite{AGT1}-\cite{AGTl},
originally motivated by the study of 5-brane compactifications
on Riemann surfaces \cite{5B1}-\cite{5Bl}.
The AGT conjecture turned the Nekrasov functions into a very serious candidate
on the role of the first special function of the string theory
\cite{AMMsf} and suggested that they indeed generalize the hypergeometric
series $\phantom._pF_q$ in the right direction.
The AGT conjecture also attracted an attention to the old problem of
conformal blocks description through the Dotsenko-Fateev integrals \cite{DF},
which are the crucial element of the free-field description of
arbitrary conformal models \cite{GMMOS,GMM}.

$\bullet$
The Dotsenko-Fateev integrals \cite{DF},
the correlators of free fields with insertion of the "screening charges",
\be
\left< \prod_a e^{i\alpha_a\phi(q_a)}
\prod_{i=1}^N \oint_{C_i} e^{ib\phi(z_i)}dz_i \right>
= \prod_{a<b}(q_a-q_b)^{2\alpha_a\alpha_b/\varepsilon^2}
\oint_{C_i} dz_i \prod_{i,a}(z_i-q_a)^{2b\alpha_a/\varepsilon^2}
\prod_{i<j}(z_i-z_j)^{2b^2/\varepsilon^2}
\label{DFI}
\ee
were long ago put into the context of
matrix models \cite{MMMconf}-\cite{discmamo} and are associated with what is
called conformal or quiver \cite{quivmamo} matrix models.
They are basically defined by the r.h.s. of (\ref{DFI}).
Then, the AGT conjecture immediately implies that these models have a direct
relation to Nekrasov functions, as anticipated in \cite{KMT},
carefully formulated in \cite{DVagt} and further discussed
in profound papers \cite{Ito,Egu}.

$\bullet$
These papers, however, stopped short from formulating the exact
shape of the Seiberg-Witten equations for the exact $\epsilon$-deformed
prepotential. The problem is to find the shape of exact SW differential $dS$,
which is very well known for $\epsilon_1=\epsilon_2=0$,
more or less understood (though still at conjectural level, see
\cite{NS,MMBS1}) for $\epsilon_2=0$ and remained a mystery for
generic $\epsilon_1$ and $\epsilon_2$.
Matrix models considerations provide the answer: $dS$ is the DV
differential $ydz$ plus {\rm higher\ genus\ corrections}, with
\be
y^2 = W'(z)^2 +4\beta f^{(0)}(z)
\ee
for the conformal matrix model potential
\be
W(z) = \sum_a^k \frac{b\alpha_a}{\varepsilon^2}\log(z-q_a)
\label{pot}
\ee
and $\beta = b^2/\varepsilon^2$.
In the simplest case of the $SU(2)$ 4-point conformal block
$k=4$, $q_{1,2,3,4} = 0,1,q_{UV},\infty$.
The higher genus corrections are described by the theory
of Dijkgraaf-Vafa phases in matrix models, \cite{DV1}-\cite{DVl},
which remains underdeveloped, but is still capable to provide
some explicit checks, see \cite{Ito,Egu}.

Relations between various parameters are summarized in the following list:
\be
\varepsilon^2 = -\epsilon_1\epsilon_2,
\ee
The central charge of the related conformal model is
\be
c = 1-\frac{6\epsilon^2}{\varepsilon^2}
= 1 - 6\left(\sqrt{\beta}-\frac{1}{\sqrt{\beta}}\right)^2,
\ \ \ \ {\rm i.e.} \ \ \ \
\epsilon = \epsilon_1+\epsilon_2 =
\varepsilon\left(\sqrt{\beta}-\frac{1}{\sqrt{\beta}}\right)
= b - \frac{\varepsilon^2}{b}
\ee
so that the screening charge is simply $b = \epsilon_1$,
another screening would have $b'=\epsilon_2=\varepsilon^2/b$,
but it does not appear in our considerations.
The conformal dimension of the vertex operator $e^{i\alpha\phi}$ is
\be
\Delta_\alpha = \frac{\alpha(\epsilon-\alpha)}{\epsilon_1\epsilon_2}
= \frac{\alpha(\alpha-\epsilon)}{\varepsilon^2}
\ee

\bigskip

\begin{figure}[t]
\begin{center}
\includegraphics[width=250pt]{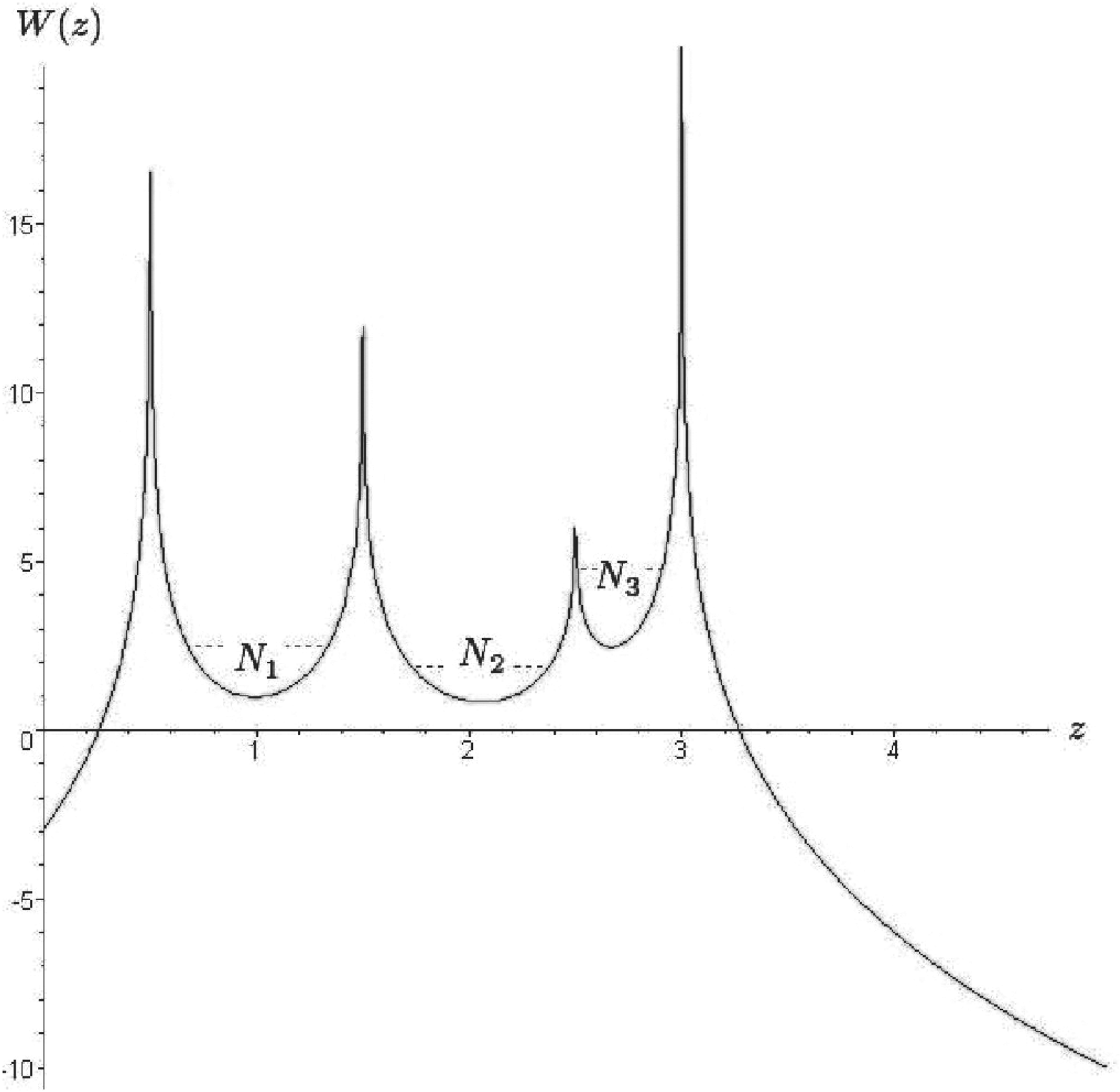}
\end{center}
\caption{Potential $-W(z)$ for four masses ${\vec \alpha} = (3,2,1,3)$
situated at positions ${\vec q} = (0.5, 1.5, 2.5, 3)$. The extrema of $W(z)$ are filled
with the eigenvalues with filling fractions $N_1, N_2, N_3$.}
\end{figure}

\begin{figure}[t]
\begin{center}
\includegraphics[width=150pt]{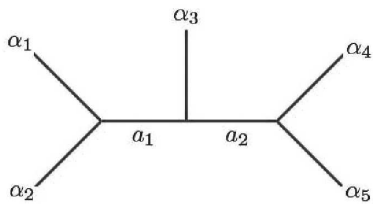}
\end{center}
\caption{Diagram for the five-point conformal block, with the dimensions of external
states corresponding to the parameters $\alpha$ and the dimensions of intermediate states
corresponding to the parameters $a_1 = N_1 - N_2$, $a_2 = N_2 - N_3$.}
\end{figure}

{\bf 3. DV phase of the $\beta$-ensemble.}
Of all these subjects the central one for this paper is explicit
formula (\ref{dS}) for exact $dS$, and it involves two advanced
notions in matrix model theory: the $\beta$-ensembles and the DV phases.
The $\beta$-ensembles of which the conformal matrix models and
Dotsenko-Fateev integrals are examples, are not, strictly
speaking, matrix models:
when the Van-der-Monde determinant $\Delta(z) = \prod_{i<j}(z_i-z_j)$
is raised to the power $2\beta$ with $\beta\neq 1,1/2,2$
they are rather not matrix but eigenvalue integrals.
However, the theory of matrix models \cite{UFN3} does not
really have so much to do with literally matrix integrals: it is rather
a theory of $\tau$-functions, subjected to additional constraints,
like string equations and, as a consequence, the set of Virasoro
or $W$-like constraints \cite{Wirc}, also known as loop equations
\cite{loops}. These equations are not very much affected by
the change of $\beta$, and all the "matrix-model" multi-resolvents
$\rho^{(p|m)}_\beta$ \cite{AMMsf} (or their Fourier/Laplace transforms,
relevant in some combinatorial contexts \cite{Ok,AMMP}) are equally well
defined for arbitrary $\beta$ (see \cite{DiF} for one of the first
reviews of the subject and \cite{Zab} for a fresh look).
In principle, for $\beta\neq 1$ the terms with half-integer $p$
(associated with non-orientable Riemann surfaces) could arise in the
large $N$ genus
expansion for arbitrary potentials \cite{DVagt,itoi},
but this does not necessarily happen in expansion (\ref{dS}).

Another ingredient of our description of exact $dS$ is the
Dijkgraaf-Vafa (DV) phase: a peculiar phase of arbitrary matrix model,
describing a multi-cut solution to the Virasoro like constraints
where the cuts are, as usual, associated with extrema of the action,
and the numbers $N_i$ of eigenvalues (filling numbers),
ascribed to each extremum are kept
fixed. This phase is well defined for all values of $N_i$, not
obligatory infinite \cite{AMMsf}, and it is characterized by the peculiar DV
differential $ydz$, which is essentially the 1-point resolvent at
genus zero $\rho^{(0|1)}(z)$. By itself, it provides the non-deformed,
original SW differential. Switching on the $\epsilon_1=-\epsilon_2$
deformation modifies it by higher genus corrections, while the
orthogonal deformation to $\epsilon = \epsilon_1+\epsilon_2\neq 0$
corresponds to making $\beta \neq 1$.

The need for the DV phase appears because the number of parameters in
the Dotsenko-Fateev integral (\ref{DFI}) does not match the number
of parameters in the conformal blocks.
The conformal block depends on dimensions of external legs: these
are labeled by parameters $\alpha_a$ in (\ref{DFI}), on the central
charge: this is hidden in $b$ (or $\beta$ or $\epsilon_1/\epsilon_2$),
and on dimensions of intermediate states: there is nothing
obvious to encode them in (\ref{DFI}).
This mismatch is the long-standing puzzle in conformal field theory.
The way out, implied in (\ref{dS}) and suggested, as we understand,
in \cite{DVagt} is that the DV phase is additionally labeled by parameters
$N_i$ (filling numbers), and these are the ones that are missed hidden parameters
in (\ref{DFI}) and they can be used to account for intermediate dimensions
in the conformal block.
The number of the extra parameters is the number of extrema of the potential
$W(z)$ in (\ref{pot}), and it is exactly the number of intermediate
states in the conformal block.
Moreover, the intermediate $\alpha$-parameters are {\it linear combinations} of
the filling factors:
\be
a_I = \oint_{A_I} dS_{\epsilon_1,\epsilon_2} \sim N_{I+1}-N_I
\ee

There are powerful techniques in matrix model theory to deal
with DV phases, based on Givental decomposition \cite{Givdeco}
on the formalism of check-operators \cite{AMMcheck},
and on Whitham dynamics \cite{ItoDV}.
Here we do not go into details of these approaches and just sketch
two of the possible lines of reasoning, which seem most promising.

\bigskip

{\bf 4. Direct description of DV phase: $\beta=1$.}
As a first illustration, we consider a particular version of the Givental
decomposition method \cite{Givdeco,KMT}.
In this particular form it is applicable only for $\beta=1$,
but it highlights essential properties of the DV phase.
For $\beta=1$ the conformal matrix model (\ref{DFI})
possesses the simple determinant representation
\be
\prod_{i=1}^n \oint_{C_i}dz_i
\prod_{i,a} (z_i-q_a)^{\alpha_a} \prod_{i<j}(z_i-z_j)^{2} = \det_{i,j} H_{ij}
\label{Nintegrals}
\ee
which admits a straightforward analytical continuation to large differences
$N_a-N_b$ without any reference to the spectral curve formalism.
Here $H_{ij}$ is the so-called matrix of moments:
\be
H_{ij}(C) = \int\limits_{C} \ dz \ z^{i + j} \prod\limits_{a} (z - q_a)^{\alpha_a}
\ee
In the DV phase there is a finite set of allowed contours, in one-to-one
correspondence with extrema of the potential.
One substitutes into (\ref{Nintegrals}) a linear combination
\be
H_{ij} = \sum_I u_I H_{ij}(C_I)
\ee
and picks up the coefficient in front of $\prod_I u_I^{N_I}$ in $\det H$, this
defines the DV phase partition function. For example, in the particular case of
the four-point function
the $H_{ij}(C)$ is the Euler hypergeometric integral
\be
H_{ij}(C) = \int\limits_{C} \ d z \ z^{i + j + \alpha_1}
(z-1)^{\alpha_2} (z-q)^{\alpha_3}
\ee
which is easily expressed through hypergeometric functions:
depending on the contour, $H_{ij}$ is equal to either
\begin{align}
\int\limits_{0}^{1} \ d z \ z^{i + j + \alpha_1}
(z-1)^{\alpha_2} (z-q)^{\alpha_3} = \dfrac{(-1)^{\alpha_2 + \alpha_3}}{1+i+j} {}_{2}F_{1}
\left( \left. \begin{array}{ccc}
-\alpha_3,\ \ -\alpha_1 - \alpha_2 - \alpha_3 - 1 - i - j \\
 - \alpha_3 - \alpha_1 - i - j \end{array} \right| \ q \ \right)
\end{align}
or
\begin{align}
\int\limits_{0}^{q} \ d z \ z^{i + j + \alpha_1}
(z-1)^{\alpha_2} (z-q)^{\alpha_3} = \dfrac{(-1)^{\alpha_2 +
\alpha_3} q^{1 + i + j + \alpha_3 + \alpha_1}}{1+i+j} \ {}_{2}F_{1}
\left( \left. \begin{array}{ccc} -\alpha_2,\ \ \alpha_1 + i + j + 1 \\
2 + \alpha_3 + \alpha_1 + i + j \end{array} \right| \ q \ \right)
\end{align}
The DV phase partition function is given by
\begin{align}
\oint\dfrac{du}{u^{N_1+1}} \oint\dfrac{dv}{v^{N_2+1}}
\det\limits_{i,j} \left( u \int\limits_{0}^{1} \ d z \ z^{i + j + \alpha_1}
(z-1)^{\alpha_2} (z-q)^{\alpha_3} + v \int\limits_{0}^{1} \ d z \ z^{i + j + \alpha_1}
(z-1)^{\alpha_2} (z-q)^{\alpha_3} \right)
\label{DVIntegral}
\end{align}
Note that the indices $i$ and $j$ run from $0$ to $N_1+N_2-1$.
This expression allows a straightforward analytical continuation to large $N_1-N_2$
in each order in $q_{UV}$.
The explicit calculation for arbitrary $\alpha_1, \alpha_2, \alpha_3$ is rather
lengthy and will be presented elsewhere.
In this paper we consider only the case of $\alpha_1 = \alpha_2 = \alpha_3 = 0$,
to provide a simple and clear evidence of relation between integral (\ref{DVIntegral})
and the 4-point conformal block.

In the case of vanishing $\alpha_i$, the partition function takes the form
\begin{align}
Z_{N_1,N_2}(q) = \oint\dfrac{du}{u^{N_1+1}} \oint\dfrac{dv}{v^{N_2+1}} \det\limits_{i,j} \left( \dfrac{u}{i + j + 1} +  \dfrac{v \ q^{i + j + 1}}{i + j + 1}  \right)
\label{DVIntegralZeroMasses}
\end{align}
From dimension counting, the partition function behaves as
\begin{align}
Z_{N_1,N_2}(q) = {\rm const}(N_1,N_2) \cdot q^{N_2^2} \cdot
\Big( C_0(N_1,N_2) + C_1(N_1,N_2) q + C_2(N_1,N_2) q^2 + \ldots \Big)
\end{align}
Thus, the quantities of interest are the normalization constant
\begin{align}
{\rm const}(N_1,N_2) = F_{N_1}(2N_2) \ F_{N_2}(0), \nn \\
F_N(\alpha) = \dfrac{1}{8 \pi} \left( \dfrac{ \alpha + N }{ N } \right)^N
\cdot \prod\limits_{L = 1}^{N-1} 16^L \ \dfrac{\Gamma\big(L+1/2\big)
\Gamma\big(L-1/2\big)}{\Gamma\big(L\big)\Gamma\big(L+1\big)}
\left( \dfrac{ (\alpha + L)(\alpha + 2 N - L) }{ L (2 N - L) } \right)^L
\end{align}
and the expansion coefficients
\begin{align}
C_0(N_1,N_2) = 1, \nn \\
C_1(N_1,N_2) = \emph{} - \dfrac{N_1^2 + 2 N_1 N_2}{2}, \nn \\
C_2(N_1,N_2) = \dfrac{ 1 - 3N_2^2 + 8N_2^4}{4(1 - 4N_2^2)^2}
\Big( N_1^2 + 2 N_1 N_2 \Big)^2  - \dfrac{ 1 - 3N_2^2}{4(1 - 4N_2^2)}
\Big( N_1^2 + 2 N_1 N_2 \Big),\nn\\
\ldots
\end{align}
Note that these coefficients are not symmetric w.r.t. $N_1$ and $N_2$.
Actually, there are no reasons to expect such a symmetry: the integration
contours of the first and of the second type are different. If one would integrate from
$0$ to $q_i$ with $i = 1,2$, then the answer would be symmetric with respect to changing
$(N_1,q_1)\leftrightarrow (N_2,q_2)$. This symmetry gets broken, when one chooses $q_1 = 1, q_2 = q$.

One observes that the poles can emerge in the $N$-parameteres,
in fact they are closely related to the t'Hooft-De Wit
spurious poles in the theory of unitary matrix integrals
\cite{tdv,unint}.
These coefficients need to be compared with the first two coefficients
of the four-point conformal block \cite{MMMagt}
\be
{\cal B}^{(0)}(\Delta_1,\Delta_2,\Delta_3,\Delta_4,\Delta)=1\nn
\ee
\be
{\cal B}_\Delta^{(1)}(\Delta_1,\Delta_2,\Delta_3,\Delta_4,\Delta) =
{(\Delta+\Delta_1 -\Delta_2   )(\Delta+\Delta_3-\Delta_4)
\over 2\Delta}
\nn
\ee
\be
{\cal B}_\Delta^{(2)}(\Delta_1,\Delta_2,\Delta_3,\Delta_4,\Delta)=
{(\Delta+\Delta_1 -\Delta_2   )(\Delta+\Delta_1 -\Delta_2   +1)
(\Delta+\Delta_3-\Delta_4)
(\Delta+\Delta_3-\Delta_4+1)\over 4\Delta(2\Delta+1)}+\\
+{\left[(\Delta_2   +\Delta_1 )(2\Delta+1)+\Delta(\Delta-1)
-3(\Delta_2   -\Delta_1 )^2\right]
\left[(\Delta_3+\Delta_4)(2\Delta+1)+\Delta(\Delta-1)
-3(\Delta_3-\Delta_4)^2\right]
\over 2(2\Delta+1)\Big(2\Delta(8\Delta-5) + (2\Delta+1)c\Big)}
\nn
\ee
On the matrix model side, putting $N_1 = n, N_2 = -n$, one finds
\be
C_0(n, - n) = 1, \nn \\
C_1(n, -n) = \dfrac{n^2}{2}, \nn \\
C_2(n,-n) = \dfrac{ n^2( 1 - 6 n^2 + 9 n^4 + 8 n^6 )}{4(1 - 4n^2)^2}
\ee
On the conformal block side this corresponds to putting
$\Delta_1 = \Delta_2 = \Delta_3 = \Delta_4 = 0$, when
\be
{\cal B}^{(0)}(0,0,0,0,\Delta)=1, \nn \\
{\cal B}_\Delta^{(1)}(0,0,0,0,\Delta) = \dfrac{\Delta}{2}\nn
\\
{\cal B}_\Delta^{(2)}(0,0,0,0,\Delta)=
\dfrac{ \Delta( 1 - 6 \Delta + 9 \Delta^2 + 8 \Delta^3 )}{4(1 - 4\Delta^2)^2}
\ee
One can see that
\begin{equation}
\addtolength{\fboxsep}{5pt}
\boxed{
\begin{gathered}
C_i(n,-n) = {\cal B}_\Delta^{(i)}(0,0,0,0,n^2), \ \ \ \ \ i = 0,1,2,\ldots
\end{gathered}
}\label{CoefficientEquality}
\end{equation}
i.e. the perturbative expansion of the matrix model
indeed reproduces correctly the perturbative expansion of the conformal block.

Moreover, this observation can be generalized: on the matrix model side,
putting $N_1 = n + \delta, N_2 = -n$, one finds
\begin{align}
C_0(n + \delta, -n) = 1, \nn \\
C_1(n + \delta, -n) = \dfrac{n^2-\delta^2}{2}, \nn \\
C_2(n + \delta, -n) = \dfrac{ (n^2-\delta^2)(-6 n^2 - \delta^2 + 9 n^4
+ 3 n^2 \delta^2 + 8 n^6 - 8 n^4 \delta^2 + 1) }{4(1 - 4n^2)^2}
\end{align}
On the conformal block side, one puts $\Delta_1 = \Delta_2 = \Delta_3 = 0$ and gets
\be
{\cal B}^{(0)}(0,0,0,\Delta_4,\Delta)=1, \nn \\
{\cal B}_\Delta^{(1)}(0,0,0,\Delta_4,\Delta) = \dfrac{\Delta - \Delta_4}{2}
\label{B1X}\nn\\
{\cal B}_\Delta^{(2)}(0,0,0,\Delta_4,\Delta)=
\dfrac{ (\Delta - \Delta_4) (- 8 \Delta^2 \Delta_4 - \Delta_4 + 3 \Delta_4 \Delta +
8 \Delta^3 + 9 \Delta^2 + 1 - 6 \Delta)}{4(1 - 4\Delta^2)^2}
\label{B2X}
\ee
Again
\begin{equation}
\addtolength{\fboxsep}{5pt}
\boxed{
\begin{gathered}
C_i(n + \delta, -n) = {\cal B}_\Delta^{(i)}(0,0,0,\delta^2,n^2), \ \ \ \ \ i = 0,1,2,\ldots
\end{gathered}
}\label{CoefficientEquality2}
\end{equation}
This two-parametric equality provides more evidence of
the equivalence between conformal block and matrix integral.
Note that in this example $N_1-N_2=2n+\delta$ is not proportional to $a=n$, the additional
constant $\delta/2$ being removable by a proper choice of the contours.
Two possible and natural generalizations are in order: to non-vanishing
$\alpha_1, \alpha_2$ and $\alpha_3$ (they should match the missing parameters
$\Delta_1, \Delta_2$ and $\Delta_3$ on the conformal block side) and to $\beta \neq 1$.
We now proceed to the second one.

\bigskip

{\bf 5. Direct description of DV phase: $\beta \neq 1$.}
For $\beta \neq 1$ and vanishing masses the partition function of the conformal matrix model
takes the form
\begin{align}
Z_{N_1,N_2}(\beta,q) = \prod\limits_{i = 1}^{N_1} \int\limits_{0}^{1} du_i \
\prod\limits_{i = 1}^{N_2} \int\limits_{0}^{q} d v_i \ \prod_{i<j}(u_i-u_j)^{2\beta} \
\prod_{i<j}(v_i-v_j)^{2\beta} \ \prod\limits_{i = 1}^{N_1} \prod\limits_{j = 1}^{N_2}
(u_i-v_j)^{2\beta}
\end{align}
No determinant representations are available for this partition function; however, the
integrand is just a polynomial and can be integrated explicitly for each particular $N_1, N_2$
and $\beta$. From dimension counting
\begin{align}
Z^{(\beta)}_{N_1,N_2}(q) = {\rm const}^{(\beta)}(N_1,N_2) \cdot q^{\beta N_2(N_2-1) + N_2}
\cdot \Big( C^{(\beta)}_0(N_1,N_2) + C^{(\beta)}_1(N_1,N_2) q +
C_2^{(\beta)}(N_1,N_2) q^2 + \ldots \Big)
\end{align}
Again, the quantities of interest are the normalization constant and the expansion coefficients.
After some calculation, one finds the normalization constant
\begin{align}
{\rm const}^{(\beta)}(N_1,N_2) = F_{N_1}^{(\beta)}(2 \beta N_2) \ F^{(\beta)}_{N_2}(0)
\end{align}
where
\begin{align}
F_N^{(\beta)}(\alpha) = G_N^{(\beta)} \cdot
\left(\dfrac{\alpha + \beta N + 1 - \beta}{\beta N + 1 -
\beta}\right)^N \cdot \prod\limits_{L = 1}^{N-1}
\left( \dfrac{\big( \beta L + 1 \big)_{\alpha}}{\big( \beta L + 1 - \beta \big)_{\alpha}}
\dfrac{\big( 2 \beta N - \beta + 2 - \beta L \big)_{\alpha}}{\big( 2 \beta N - 2 \beta + 2 -
\beta L \big)_{\alpha}} \right)^L
\end{align}
and $(a)_k = \Gamma(a+k)/\Gamma(a)$. We are yet unable to find $G_N^{(\beta)}$ explicitly; only
for $\beta = 1$ we know that
\begin{align}
G_N^{(1)} = \dfrac{1}{8 \pi} \prod\limits_{L = 1}^{N} \ 16^L \ \dfrac{\Gamma\big(L+1/2\big)
\Gamma\big(L-1/2\big)}{\Gamma\big(L\big)\Gamma\big(L+1\big)}
\end{align}
Also one finds the coefficients
\begin{align}
C_0(N_1,N_2) = 1
\end{align}
\begin{align}
C_1(N_1,N_2) = \emph{} - \beta \dfrac{N_1^2 + 2 N_1 N_2}{2} + \dfrac{\beta - 1}{2} N_1
\end{align}
Comparison with the corresponding coefficient in the conformal block gives
\begin{align}
\boxed{
C_1(n,-n) = \beta \dfrac{n^2}{2} + \dfrac{\beta - 1}{2} n = \dfrac{\Delta}{2} =
{\cal B}_\Delta^{(1)}(0,0,0,0,\Delta)
}\label{b1}
\end{align}
This is in accordance with the formula
\begin{align}
\Delta = \dfrac{\alpha( \epsilon_1 + \epsilon_2 - \alpha)}{\epsilon_1 \epsilon_2}
\end{align}
Indeed, on dimensional grounds $\alpha = A n$ and one obtains
\begin{align}
\dfrac{A n( \epsilon_1 + \epsilon_2 - A n)}{\epsilon_1 \epsilon_2} = \beta \dfrac{n^2}{2} +
\dfrac{\beta - 1}{2} n
\end{align}
A system of two equations
\begin{align}
A \dfrac{\epsilon_1 + \epsilon_2}{2 \epsilon_1 \epsilon_2} = \dfrac{\beta - 1}{2}, \ \ \ \
\dfrac{A^2}{2 \epsilon_1\epsilon_2} = -\dfrac{\beta}{2}
\end{align}
is satisfied by $A = -\epsilon_1, \ \beta = -\epsilon_1/\epsilon_2$ (or $A =
-\epsilon_2, \ \beta = -\epsilon_2/\epsilon_1$).
Thus, we derived "from the first principles" the
Dotsenko-Fateev integral representation of the conformal block.

Like in the case of $\beta=1$ eq.(\ref{b1}) can be easily extended to include $\Delta_4\ne 0$:
\begin{align}
\boxed{
C_1(n+\delta,-n) =
{\cal B}_\Delta^{(1)}(0,0,0,\Delta_4,\Delta)
}
\end{align}
Because of our choice of contours we have $\alpha=-\epsilon_1 (n+2\delta)$,
$\alpha_4=-\epsilon_1\delta$.

\bigskip

{\bf 6. Genus expansion and loop equations.}
We now turn to a more traditional approach to the genus expansion
\cite{AMMsf},
straightforwardly applicable for any values of $\beta$.
In the case of DV phases it has certain peculiarities
(the formalism of check-operators \cite{AMMcheck} has to be very effective
in this case).

The Ward identities for the conformal matrix model follow as usual
\cite{Wirc} from
invariance under the shift of integration variables
$\delta z_i \sim z_i^{n+1}$ with $n\geq -1$.
Invariance of the r.h.s. of (\ref{DFI}) implies
\be
\delta\Big(\!\!<1>\!\!\Big) = \delta\left(
\prod_{i=1}^N \oint dz_i \prod_{a}(z_i-q_a)^{\mu_a/\varepsilon^2}
\prod_{i<j}^N(z_i-z_j)^{2\beta}\right)
=  0,
\ee
i.e.
\be
\left< 2\beta\varepsilon^2\sum_{i<j}\frac{z_i^{n+1}-z_j^{n+1}}{z_i-z_j}
+ \sum_i\sum_a \frac{\mu_az_i^{n+1}}{z_i-q_a}
+ \varepsilon^2 \sum_i nz_i^n\right>
\ = 0
\ee
where $\mu_a = 2b\alpha_a = 2\epsilon_1\alpha_a$.
Summing up these identities over $n$ with the weights $\xi^{-n-2}$,
one obtains the Virasoro constraints in the form of a loop equation
\be
\left<\beta\varepsilon^2\left(\sum_{i}\frac{1}{(\xi-z_i)}\right)^2
+ \sum_a \frac{\mu_a}{\xi-q_a}\sum_i\left(\frac{1}{\xi-z_i} +
\frac{1}{z_i-q_a}\right)
+ (1-\beta)\varepsilon^2 \sum_i \frac{1}{(\xi-z_i)^2}\right> \ = 0
\label{WI1}
\ee
In (\ref{dS}) one needs the resolvent  \cite{AMMsf}
\be
\rho_\beta^{(\cdot|1)}(\xi) = \varepsilon^2
\left<\sum_{i=1}^N\frac{1}{\xi - z_i}\right>
= \sum_{p=0}^\infty \varepsilon^{2p}\rho_\beta^{(p|1)}(\xi)
\ee
which satisfies the equation
\be
\boxed{
\beta \left(\rho^{(\cdot|1)}(\xi)\right)^2 + W'(\xi)\rho^{(\cdot|1)}(\xi) -
f(\xi) = (1-\beta)\varepsilon^2 \frac{\p \rho^{(\cdot|1)}(\xi)}{\p\xi}-
\beta\varepsilon^2\rho^{(\cdot|2)}(\xi,\xi)
}
\label{gen}
\ee
where $W'(\xi) = \sum_a \frac{\mu_a}{\xi-q_a}$
\be
\rho_\beta^{(\cdot|2)}(\xi_1,\xi_2) = \varepsilon^2
\left<\sum_{i=1,j}^N\frac{1}{(\xi_1 - z_i)(\xi_2 - z_j)}\right>_c
\ee
and
\be
f(\xi) = \sum_a \frac{\mu_a}{\xi-q_a}
\left<\sum_{i=1}^N\frac{1}{q_a-z_i}\right> =
\sum_a \frac{c_a}{\xi - q_a}
\label{f}
\ee
$<\ldots>_c$ here denotes the connected correlator, and the constants $c_a$
can {\it not} be defined from the
symmetry considerations (from the Ward identities).

Eq.(\ref{gen}) does not define the one-point resolvent directly:
the best one can do at once is to find the spherical term with $p=0$,
when the first term in (\ref{WI1}) factorizes:
\be
\boxed{
\beta \left(\rho^{(0|1)}(\xi)\right)^2 + W'(\xi)\rho^{(0|1)}(\xi) -
f^{(0)}(\xi) = 0
}
\label{gen0}
\ee
where
\be
f^{(0)}(\xi) = \sum_a \frac{\mu_a}{\xi-q_a}
\left<\sum_{i=1}^N\frac{1}{q_a-z_i}\right>_0 =
\sum_a \frac{c_a^{(0)}}{\xi - q_a}
\label{f0}
\ee
After rescaling of $\rho^{(0|1)}$ the loop equation (\ref{gen0}) can be interpreted in
terms of average of the stress tensor \cite{loops,MMMP,HDV,AGT1,Ito},
the term at the r.h.s. corresponds to adding the
$\left(\sqrt{\beta}-\frac{1}{\sqrt{\beta}}\right)\p^2\phi$
in the case of non-unit central charge.

Now solving the quadratic equation (\ref{gen0}),
one gets \cite{Egu}:
\be
\rho^{(0|1)}(\xi) = \frac{-W'(\xi) + y(\xi)}{2\beta},\ \ \ \ \ \ \ \
y^2 = W'(\xi)^2 + 4\beta f^{(0)}(\xi)
\label{rho01}
\ee
Neglecting the total derivative in (\ref{rho01}), one obtains that
\be
dS(\beta=1) \sim  y(z)dz
\label{dS0}
\ee
This SW differential has poles at the points $q_a$ and also at
$\xi=\infty$, and according to the general rules of \cite{SW2}
residues at poles are the masses of matter supermultiplets
($4=2N_C$ fundamental and $k-3$ bifundamental for $k+1>4$).
These masses are linear in $\alpha_a$ as a part of
the AGT relation \cite{AGT1}.

In the simplest $U(2)$ case of the 4-point conformal block,
i.e. for $k+1=4$,
and for all external   $\alpha_{1,2,3,4}=0$ (thus $W'(z) = 0$)
and also for $\epsilon=0$
\be
ydz = \frac{\sqrt{u}dz}{\sqrt{z(z-1)(z-q_{UV})}}
\label{zeroma}
\ee
Note that $c$-parameters in (\ref{f0}) are no obliged to vanish
together with $\mu_a$, despite they can seem to do so from that
formula: the possibility to treat $c$ as completely free parameters
and take the limits of this kind is the specifics of DV phases.
The modulus $a\sim\sqrt{u}\sim N_1-N_2$.
For $ydz$ in the case of $W(z)\neq 0$ see \cite{Egu},
in the inverse limit of pure gauge theory
$ydz = \frac{dz}{z}\sqrt{u - z- \frac{1}{z}\ }$\
becomes the sine-Gordon presymplectic form \cite{GKMMM}.
In the zeroth order in $\varepsilon$
one obtains the classical contribution
to the prepotential:
$
{\cal F} = \tau a^2 + O(\varepsilon),
$
with
\be
\tau = \frac{1}{2\pi i}\log q
= \frac{\int_0^{q_{UV}} \frac{dz}{ \sqrt{z(z-1)(z-q_{UV})} }}
{\int_0^1 \frac{dz}{\sqrt{z(z-1)(z-q_{UV})}}  } \ \
\Longrightarrow \ \
q_{UV} = \frac{\theta_{10}^4(\tau)}{\theta_{00}^4(\tau)}
\ee
which gives the celebrated relation between the bare $q_{UV}$ and
the instanton-corrected dressed coupling $q$ \cite{Mar,AGT1,Zamlim}.
This should be compared with the expression through the Nekrasov functions,
\be
{\cal F}_{Nek}(\varepsilon=0) = \lim_{\epsilon_2\rightarrow 0}
\epsilon_1\epsilon_2\log
\left(1 +
\frac{q_{UV}}{\epsilon_1\epsilon_2}\frac{a^4}{4a^2-\epsilon_1^2} + O(q_{UV}^2)
\right)
= q_{UV}\frac{a^4}{4a^2-\epsilon_1^2} + \ldots =
q_{UV}a^2 + \frac{1}{4}q_{UV}\epsilon_1^2 + \ldots
\label{nekexpan}
\ee
In the case of finite $\beta$ parameter $\epsilon^2 = \beta\varepsilon^2+
O(\varepsilon^4)$, and only the first term here survives in the genus zero
approximation. Thus, indeed the term with $\epsilon=0$ in (\ref{nekexpan})
is immediately reproduced from our consideration.

\bigskip

{\bf 7. $\epsilon_2=0$ and other limits.}
The interesting limit $\epsilon_2=0$ with finite $\epsilon_1$ was considered in much more
detail in \cite{MMBS1}.
This corresponds to keeping $\varepsilon=0$, while
$\beta\sim \epsilon_1^2/\varepsilon^2$.
In this case the r.h.s. in (\ref{gen}) can not be neglected,
instead one gets a typical Ricatti equation for $\rho^{(0|1)}$,
which should be compared with
\be
\left(-\epsilon_1^2\p^2 + e^{ix}+e^{-ix} - u\right)\psi = 0
\ \ \
\stackrel{\psi(x) = \exp\left(\frac{i}{\epsilon_1}\int^x p dx\right)}{\Longrightarrow}
\ \ \ p^2 + \epsilon_1 p' = V(z)
\ee
of \cite{MMBS1}.
In the case of the pure gauge $U(2)$ theory $V(z) = u -z-\frac{1}{z}$, $z=e^{ix}$
is the sine-Gordon potential, while the case
$N_f\neq 0$ is described in terms of XXX (at $d=4$), XXZ (at $d=5$)
and XYZ (at $d=6$) magnetics \cite{GMMMmag,GGM,MarMir},
and for $\epsilon_1\neq 0$ the counterpart
of Shroedinger like equations of \cite{MMBS1} is a somewhat more
complicated Baxter equation
\be
\left(K(-i\epsilon_1\p) + e^{ix/2}K_+(-i\epsilon_1\p)e^{ix/2}
+ e^{-ix/2}K_-(-i\epsilon_1\p)e^{-ix/2} \right)\psi = 0
\label{BaxXXX}
\ee
with $K_{\pm}(p) = \prod_{a=1}^{N_c} (p - m_a^{\pm})$.
This is a "quantization" of the spectral curve \cite{GMMMmag}
$K(p) = z + \frac{Q(p)}{z}$ with the SW differential $dS=p\frac{dz}{z}$.
where $z = e^{ix}K_+(p)$ and $Q(p) = K_+(p)K_-(p)$.
Even more sophisticated equations are needed in the case of toric $1$-point
function, which is associated with adjoint matter and where one needs a
quantum deformation of the Calogero system. In the $U(2)$ case it is just given by the
Weierstrass function potential $V(x)=g\wp(x)$, while at higher $U(n)$ one has to work
with the equation in separated variables. In particular, for $U(3)$  \cite{Calsv}
\be
i\psi'''+u_1\psi''-i\Big(u_2+3g(g-1)\wp(x)\Big)\psi'-
\Big(u_3+g(g-1)\wp(x)u_1-ig(g-1)(g-2)\wp'(x)
\Big)\psi=0
\ee
$u_i$ here are the conserved quantities and $g$ is the Calogero coupling constant.

More detailed comparison of (\ref{dS}) at $\epsilon_2=0$
with \cite{MMBS1} is rather straightforward,
however, one should take into account that
(\ref{dS0}) is not the conventional
form of the SW differential for $N_f=2N_c$ implied by the
Lax formalism for the XXX magnet \cite{GMMMmag}.
In particular, in (\ref{dS0}) it is not quite simple
to take the limit to the pure gauge theory, where \cite{MMBS1}
provides a nice description.

It is instructive to study various other limits of formula (\ref{dS}).
In fact, the entire variety of limits arising in AGT theory
is of interest: that of large $\alpha_{1,2,3,4}$, leading from $N_f=2N_c$
case to the pure gauge theory $N_f=0$ \cite{puregauge};
that of large intermediate dimension
$a \sim N_1-N_2$, associated with Zamolodchikov's asymptotic \cite{Zamas}
of conformal blocks \cite{Zamlim};
that of the large central charge $c$, giving rise
to the ordinary hypergeometric series \cite{largec};
and the already mentioned limit of $\epsilon_2=0$ \cite{NS,MMBS1},
where the exact BS periods are solutions to the Baxter equations.
Unfortunately, all these limits are not so easy to take, and each one deserves
a separate discussion.
It deserves emphasizing that study of the DV phases is non-trivial.
In the literature there is something known about the prepotential
expansion in positive powers of $S$-variables (this is known as
the problem of CIV potential \cite{DV1}).
What we need now is rather an expansion in negative powers of {\it differences}
between the $S$-variables.
This is a tedious calculational problem, still relatively straightforward,
as we saw above in section 5.
However, taking limits in this sophisticated procedure is an additional
non-trivial step. A check in a rather simple situation of $\beta=1$
and $N_f = 4 \longrightarrow N_f=3,2$ in \cite{Egu} is a nice illustration of
the difficulties one needs to overcome.

\bigskip

{\bf 8. Check operators.}
Higher genus corrections can be analyzed in a similar way, starting
from the full loop equation,
\be
\beta\rho(z)^2+\beta\varepsilon^2\hat\nabla(z)\rho(z)+
\Big(W'(z)+v'(z)\Big)\rho(z)
+(\beta-1)\varepsilon^2 \frac{\p \rho(z)}{\p z}
=\left[\Big(W'(z)+v'(z)\Big)\rho(z)\right]_+
\label{leq}
\ee
The auxiliary potential $v(z) = \sum_{k=0}^\infty t_kz^k$
is used to produce higher multi-resolvents by application of the operators
$\hat\nabla(z) = \sum_{k=0}^\infty \frac{1}{z^{k+1}}\frac{\p}{\p t_k}$,
see the extended review \cite{AMMsf} for details.
The projector at the r.h.s. of (\ref{leq}) is defined
\be
\left[\Big(W'(z)\Big)\rho(z)\right]_+
=\left<\sum_a \frac{W'(z)-W'(q_a)}{z-q_a}\right>\
= \sum_a \frac{c_a}{z-q_a}
\ee
A reasonable technique to deal with these kind of quantities is
provided by {\it the check operators} \cite{AMMcheck}
\be
\check R(z) = [W'(z)\hat\nabla(z)]_+ =\sum_a \frac{\mu_a}{z-q_a}\hat\nabla(q_a)
\ee
For example,
\be
\rho^{(1|1)} = -\beta\left(\frac{y''}{4y^2} + \frac{\check R y}{2y^2}\right) +
\frac{\check R F_1}{y} +{(1-\beta)\over y}{\partial\rho^{(0|1)}(z)\over\partial z}
\label{rho11}
\ee
where
\be
-\beta \frac{y''}{4y^2} = \frac{\beta}{16 y}
\left( \left(\frac{1}{z} + \frac{1}{z-1} + \frac{1}{z-q_{UV}}\right)^2
- 2 \left(\frac{1}{z^2} + \frac{1}{(z-1)^2} + \frac{1}{(z-q_{UV})^2}\right)
\right)
\ee
Note that the $A$-periods and thus the moduli $a$ are usually non-affected
by higher genus corrections, but these additional corrections can change
the situation in the present case.

\bigskip

{\bf 9. $Z_{Nek}$ as a $\tau$-function.}
The $\varepsilon$-deformation in (\ref{dS}) is in the direction of
$\epsilon_1=-\epsilon_2$, where the central charge $c=1$, thus $\beta=1$
and one deals with an ordinary matrix model.
It is well known that inclusion of contributions of all genera
in the matrix model partition function converts the quasiclassical (Whitham)
$\tau$-function in the spherical approximation into the full KP/Toda $\tau$-function
\cite{UFN3,GKLMM}.
Thus we know the meaning of one of the two deformations in (\ref{dS}).
The other deformation, in the direction of $\epsilon_1$ with $\epsilon_2=0$
is known \cite{MMBS1} to convert the quasiclassical Bohr-Sommerfeld periods in
the formulation of SW theory into the exact BS periods, thus we also know the
meaning of the other deformation.
From the point of view of integrability theory, this implies that the
$\psi$-function, of which the exact BS periods are monodromies, is the Baker-Akhiezer
function of some full (dispersionful)
integrable system (Toda, Calogero, Ruijsenaars
or magnetic) obtained by switching from the Whitham Lax representation to that in
terms of differential operators.
At the same time, the other deformation
implies that new time-variables are introduced.
Unfortunately, these two deformations are not orthogonal.
The whole story is a generalization of two-step deformation of
the equation $p^2-u=0$: first into the Shroedinger equation $\hat L\psi
\equiv
(\p^2-u)\psi = 0$
and second into the KdV equation, where $u$ becomes a solution to the additional
hierarchy of equations $\frac{\p u}{\p t_k} = (\hat L^{k+1/2})_+$.
This picture can be similar to the vision expressed in \cite{NS}.

\bigskip

{\bf 10. Conclusion.}
To conclude, we presented a complete description of Nekrasov functions
with arbitrary values of parameters $\epsilon_1, \epsilon_2$
and the AGT associated conformal blocks in terms of the exact BS periods and thus
in the original terms of SW theory.
The main ingredient is the SW differential with two deformation
parameters $\epsilon_1$ and $\epsilon_2$, and it is given by the
matrix model resolvent (\ref{dS}), analytically continued to the DV phase
and for $\beta$, which can be different from unity.
This conjecture still needs to be thoroughly examined and proved,
but there remain few doubts that it is true.
Time is also coming to transfer knowledge from one of the so unified
subjects to others.
As a simple, but still impressive illustration we explained how the Dotsenko-Fateev
integrals are now fully matched to the generic Virasoro conformal blocks.
There are many more applications of this kind to come in the close future.

\bigskip

{\bf 11. Acknowledgements.}

Our work is partly supported by Russian Federal Nuclear Energy Agency, by RFBR grants 07-02-00878
(A.Mir.), and 07-02-00645 (A.Mor. \& Sh.Sh.), by joint grants 09-02-90493-Ukr,
09-02-93105-CNRSL, 09-01-92440-CE, 09-02-91005-ANF and by Russian President's Grant of Support
for the Scientific Schools NSh-3035.2008.2.
The work of Sh.Shakirov is also supported in part by Moebius Contest Foundation for Young
Scientists.

\end{document}